\documentclass{article}

\usepackage{arxiv}

\usepackage[utf8]{inputenc} 
\usepackage[T1]{fontenc}    
\usepackage{hyperref}       
\usepackage{url}            
\usepackage{booktabs}       
\usepackage{amsfonts}       
\usepackage{nicefrac}       
\usepackage{microtype}      
\usepackage{lipsum}

\title{On the multiplicity of the martingale condition: Spontaneous symmetry breaking in Quantum Finance}

\author{
 Ivan Arraut\thanks{This work started at the University of Tokyo} \\
  Lee Shau Kee School of Business and Administration\\
 The Open University of Hong Kong,\\
  30 Good Shepherd Street, Homantin, Kowloon \\
  \texttt{ivanarraut05@gmail.com} \\
   \And
 Alan Au \\
   Lee Shau Kee School of Business and Administration\\
 The Open University of Hong Kong,\\
  30 Good Shepherd Street, Homantin, Kowloon \\
  \texttt{kmau@ouhk.edu.hk} \\
  \AND
   Alan Ching-biu Tse \\
 Department of Marketing, The Chinese University of Hong Kong, \\
  Cheng Yu Tung Building 12 Chak Cheung Street Shatin, N.T., Hong Kong \\\\
  Lee Shau Kee School of Business and Administration\\
 The Open University of Hong Kong,\\
  30 Good Shepherd Street, Homantin, Kowloon \\
 \And
}

\begin{document}
\maketitle

\begin{abstract}
We demonstrate that the martingale condition in the stock market can be interpreted as a vacuum condition when we express the financial equations in the Hamiltonian form. We then show that the symmetry under the changes of the prices is spontaneously broken in general and the symmetry under changes in the volatility, for the case of the Merton-Garman (MG) equation, is also spontaneously broken. This reproduces a vacuum degeneracy for the system. In this way, we find the conditions under which, the martingale condition can be considered to be a non-degenerate vacuum. This gives us a surprising connection between spontaneous symmetry breaking and the flow of information through the boundaries for the financial systems. Subsequently, we find an extended martingale condition for the MG equation, depending not only prices but also on the volatility and finally, we show what happens if we include additional non-derivative terms on the Black Scholes and on the MG equations, breaking then some other symmetries of the system spontaneously.       
\end{abstract}


\section{Introduction}

In probability theory, a martingale process is the one where the future expectation value of a random variable is just the present value \cite{1}. In Quantum Finance, the martingale condition corresponds to a risk-neutral evolution, consistent with the most basic concepts taken from probability theory \cite{2}. A neutral evolution, is free from any possibility of arbitrage \cite{2, 3}. Arbitrage in finance gives the possibility of investors to operate in different markets \cite{3}. For example, in general it is possible for a broker to buy sharings in New York and then sell them in Hong Kong, getting in this way, some income from the differences in the prices in both markets over the same product. This can be done in a market outside the equilibrium condition (outside martingale). In addition, only big corporations can get benefits with some income from arbitrage due to the high volume in their inversions. Individuals cannot receive enough earnings by using arbitrage due to the fee charges in transactions which would cancel any possibility of income \cite{3}. The arbitrage process helps the market to arrive to the equilibrium state. Once the market is in this equilibrium condition, any possibility or arbitrage is lost. It is at this point when we have a martingale process with a risk-neutral evolution. In fact, the existence of a martingale condition is known as {\bf the fundamental theorem of finance} \cite{3, 4}.
In this paper we demonstrate that when we express some financial equations, like the Black-Scholes (BS) or the Merton-Garman (MG) equations, in the Hamiltonian form; then the martingale condition can be interpreted as a vacuum condition \cite{3}. In the most general sense, this vacuum condition is non-unique for the BS and the MG cases when we explore the symmetries under change of prices and the symmetry under changes in volatility for the MG case. These symmetries come out to be spontaneously broken \cite{7} (they do not annihilate the vacuum state), except for some combination of parameters, which guarantee the vacuum (martingale state) to be unique ( non-degenerate). The same mentioned combination of parameters guarantee the no-flow of information through the boundaries of the system. This is an interesting connection between spontaneous symmetry breaking and flow of information, already perceived in a different context in \cite{Winchi}. Subsequently we formulate an extended martingale condition for the case of the MG equation, which considers the martingale state not only as a function of the prices of the stock, but also on the stochastic volatility. For this case again the vacuum, taken as the martingale state, is degenerate and then the corresponding symmetries are spontaneously broken. The conditions for the vacuum to be single for this case, come out to be not only the necessity of no-flow of information through the boundaries of the system, but also the absence of any white noise coming from the stochastic volatility. These two conditions guarantee the martingale state (vacuum) to be single, restoring then the symmetries for the vacuum state. Finally, we study an ideal situation where for the MG and the BS case, we add non-derivative terms, such that we still satisfy the martingale condition. Different potential terms have been analyzed before by some other authors \cite{3, 5}. In \cite{6}, the double slit constraint was considered instead of a formal potential term in the Hamiltonian in order to model sudden fluctuations in the market. Following this game,  in this paper we also investigate what happens if we add two non-derivative terms with two additional free-parameters (standard situation). We analyze the case where the combinations of the two free-parameters gives a degenerate vacuum condition, such that new symmetries are spontaneously broken. These particular situations are relevant when the kinetic terms for the BS as well as for the MG equations are negligible in the neighborhood of a well-defined vacuum. The paper is organized as follows: In Sec. (\ref{Sec1}), we describe the BS equation and we express it in its Hamiltonian form. In Sec. (\ref{MGequationsec}), we explain the MG equation and its corresponding Hamiltonian. In Sec, (\ref{Sec2}), we explain the meaning of a martingale condition from the classical perspective in Finance. In Sec. (\ref{Sec3}), we explain why the martingale condition, for the evolution of an Option, is equivalent to a vacuum condition from the perspective of Quantum Mechanics. In Sec. (\ref{Sec4}), we introduce some potential terms in the financial Hamiltonians and we analyze under which conditions they preserve the martingale condition. In Sec. (\ref{NS1}), in the scenario of the BS and the MG equation, we show that the symmetries under a changes in the prices are spontaneously broken, giving then a multiplicity of vacuums (martingale states), except for some particular combinations of the parameters of the Hamiltonian. The same is applied for the symmetries under changes of the volatility for the MG equation, when we extend the notion of martingale states in order to include the volatility as one of the variables. In Sec. (\ref{NS2}), we add some potential terms to the BS and the MG equations, introducing then additional broken symmetries in the system. The vacuum of the added potential is such that in its neighborhood, the MG and the BS equations will have the same behavior, because the kinetic terms can be neglected. Finally, in Sec. (\ref{Sec5}), we conclude.

\section{The Black-Scholes equation}   \label{Sec1}

The stock price $S(t)$ is normally taken as a random stochastic variable evolving in agreement to a stochastic differential equation given by

\begin{equation}   \label{rainbowl}
\frac{dS(t)}{dt}=\phi S(t)+\sigma SR(t).
\end{equation}   
Here $\phi$ is the expected return of the security, $R(t)$ is the Gaussian white noise with zero mean and $\sigma$ is the volatility \cite{3}. Note that this simple equation contains one derivative term on the left-hand side and non-derivative terms on the right-hand side. The fundamental analysis of Black and Scholes, exclude the volatility such that we can guarantee the evolution of the price of the stock with certainty \cite{9}. In this way, by imposing $\sigma=0$, we obtain a simple solution for the equation (\ref{rainbowl}) as

\begin{equation}
S(t)=e^{\phi t}S(0).    
\end{equation}
The possibility of arbitrage is excluded if we can make a perfect hedged portfolio. In this sense, any possibility of uncertainty is excluded and we can analyze the evolution of the price free of any white noise \cite{3}. We can consider the following portfolio

\begin{equation}   \label{forever}
\Pi=\psi-\frac{\partial \psi}{\partial S}S.
\end{equation}
This is a portfolio where an investor holds the option and then {\it short sells} the amount $\frac{\partial \psi}{\partial S}$ for the security $S$. By using the Ito calculus (stochastic calculus) \cite{3}, it is possible to demonstrate that

\begin{equation}   \label{BS}
\frac{d\Pi}{dt}=\frac{\partial \psi}{\partial t}+\frac{1}{2}\sigma^2S^2\frac{\partial^2 \psi}{\partial S^2}.
\end{equation}
Here the change in the value of $\Pi$ does not have any uncertainty associated to it \cite{3}. The random term has disappeared due to the choice of portfolio. Since here we have a risk-free rate of return for this case (no arbitrage) \cite{Europe, Epjons}, then the following equation is satisfied

\begin{equation}   \label{hedged}
\frac{d\Pi}{dt}=r\Pi.
\end{equation} 
If we use the results (\ref{forever}) and (\ref{BS}), together with the previous equation, then we get

\begin{equation}   \label{BSeq}
\frac{\partial \psi}{\partial t}+rS\frac{\partial \psi}{\partial S}+\frac{1}{2}\sigma^2S^2\frac{\partial^2\psi}{\partial S^2}=r\psi.
\end{equation}
This is the Black-Scholes equation \cite{Op2, Op3, Merton2}, which is independent of the expectations of the investors, defined by the parameter $\phi$, which appears in eq. (\ref{rainbowl}). In other words, in the Black-Scholes equation, the security (derivative) price is based on a risk-free process. The Basic Assumptions of the Black-Scholes equation are:\\
1). The spot interest rate $r$ is constant.\\
2). In order to create the hedged portfolio $\Pi$, the stock is infinitely divisible, and in addition it is possible to short sell the stock.\\
3). The portfolio satisfies the no-arbitrage condition.\\
4). The portfolio $\Pi$ can be re-balanced continuously.\\
5). There is no fee for transaction.\\ 
6). The stock price has a continuous evolution. \\

\subsection{Black-Scholes Hamiltonian formulation}

We will explain how the eq. (\ref{BSeq}), can be expressed as an eigenvalue problem after a change of variable. The resulting equation will be the Schr\"odinger equation with a non-Hermitian Hamiltonian. For eq. (\ref{BSeq}), consider the change of variable $S=e^x$, where $-\infty<x<\infty$. In this way, the BS equation becomes

\begin{equation}   \label{BSHamiltonian2}
\frac{\partial\psi}{\partial t}=\hat{H}_{BS}\psi,
\end{equation} 
where we have defined the operator  

\begin{equation}   \label{BSHamiltonian}
\hat{H}_{BS}=-\frac{\sigma^2}{2}\frac{\partial^2}{\partial x^2}+\left(\frac{1}{2}\sigma^2-r\right)\frac{\partial}{\partial x}+r.
\end{equation}
as the BS Hamiltonian. Note that the resulting Hamiltonian is non-Hermitian since $\hat{H}\neq\hat{H}^+$ \cite{5}. In addition, note that since the spot interest rate $r$ is constant, then the potential term is just a constant term. This means that the vacuum condition is trivial for this case. Under the BS Hamiltonian, the evolution in time of the Option is non-unitary in general (in addition, the Hamiltonian non-necessarily obeys the $PT$ symmetry). This means that the probability is not necessarily preserved in time, although it is certainly well-defined and its total value is equal to one. In general, there are some cases in ordinary Quantum Mechanics, as well as in Quantum Field Theory, where it is interesting to explore non-Hermitian Hamiltonians (Lagrangians) \cite{8}. Based on the previous explanations, we cannot expect the financial market to obey unitarity. The reason for this is simply because the market is not a closed system and there are many external factors influencing its behavior as for example it is the amount of people and organizations trading at some instant of time. Then the assumption of unitarity makes no sense at all and the Hamiltonian must be non-Hermitian. In this paper however, when we add some potential terms to the BS and MG equations, we will impose the Hermiticity condition on them. When the symmetry is spontaneously broken and we are working at the vacuum level, all the terms in the original BS equation become irrelevant, giving then importance to (only) the potential terms. In this way we can follow the standard formalism suggested in \cite{8}. The same conclusions apply to any other equation only containing kinetic terms as it is the case of the Merton-Garman (MG) equation to be analyzed shortly. Inside the MG case however, the conclusions about what is happening during the breaking of the symmetry of the market (spontaneously) are more relevant than in the BS case. We will return back to this point later in this paper.        

\section{The Merton-Garman equation: Preliminaries and derivation}   \label{MGequationsec}

We can consider a more general case where the security and the volatility are both stochastic. In such a case, the market is incomplete \cite{3}. Although several stochastic processes have been considered for modeling the case with stochastic volatility \cite{45}, here we consider the generic case, defined by the set of equations \cite{3}

\begin{eqnarray}   \label{corona5}
\frac{dS}{dt}=\phi Sdt+S\sqrt{V}R_1\nonumber\\
\frac{dV}{dt}=\lambda+\mu V+\zeta V^\alpha R_2.
\end{eqnarray}
Here the volatility is defined through the variable $V=\sigma^2$ and $\phi$, $\lambda$, $\mu$ and $\zeta$ are constants \cite{51}. The Gaussian noises $R_1$ and $R_2$, corresponding to each of the variables under analysis, are correlated in the following form

\begin{equation}   \label{corona3}
<R_1(t')R_1(t)>=<R_2(t')R_2(t)>=\delta(t-t')=\frac{1}{\rho}<R_1(t)R_2(t')>.    
\end{equation}
Here $-1\leq\rho\leq1$, and the bra-kets $<AB>$ correspond to the correlation between $A$ and $B$. If we consider a function $f$, depending on the Stock price, the time, as well as on the white noises; with the help of the Ito calculus, it is possible to derive the total derivative in time of this function as

\begin{eqnarray}
\frac{df}{dt}=\frac{\partial f}{\partial t}+\phi S\frac{\partial f}{\partial S}+(\lambda+\mu V)\frac{\partial f}{\partial V}+\frac{\sigma^2S^2}{2}\frac{\partial^2f}{\partial S^2}+\rho V^{1/2+\alpha}\zeta\frac{\partial^2f}{\partial S\partial V}+\frac{\zeta^2V^{2\alpha}}{2}\frac{\partial^2f}{\partial V^2}+\sigma S\frac{\partial f}{\partial S}R_1+\zeta V^\alpha\frac{\partial f}{\partial V}R_2.    
\end{eqnarray}
This equation can be expressed in a more compact form which separates the stochastic terms from the non-stochastic ones as follows

\begin{equation}   \label{HK}
\frac{df}{dt}=\Theta+\Xi R_1+\psi R_2.    
\end{equation}
 Here we have defined 
 
\begin{eqnarray}
\Xi=\sigma S\frac{\partial f}{\partial S}, \;\;\;\;\; \;\;\;\;\; \;\;\;\;\;\psi=      
\zeta V^\alpha\frac{\partial f}{\partial V},\nonumber\\
\Theta=\frac{\partial f}{\partial t}+\phi S\frac{\partial f}{\partial S}+(\lambda+\mu V)\frac{\partial f}{\partial V}+\frac{\sigma^2S^2}{2}\frac{\partial^2f}{\partial S^2}+\rho V^{1/2+\alpha}\zeta\frac{\partial^2f}{\partial S\partial V}+\frac{\zeta^2V^{2\alpha}}{2}\frac{\partial^2f}{\partial V^2},
\end{eqnarray} 
 keeping in this way the notation used in \cite{3} for convenience. 
 
 \subsection{Derivation of the Merton-Garman equation}
 
 If we consider two different options defined as $C_1$ and $C_2$ on the same underlying security with strike prices and maturities given by $K_1$, $K_2$, $T_1$ and $T_2$ respectively. It is possible to create a portfolio 
 
 \begin{equation}
\Pi=C_1+\Gamma_1C_2+\Gamma_2S.     
 \end{equation}
 If we consider the result (\ref{HK}), then we can define the total derivative with respect to time as for the folio as
 
 \begin{equation}
\frac{d\Pi}{dt}=\Theta_1+\Gamma_1\Theta_2+\Gamma_2\phi S+\left(\Xi_1+\Gamma_1\Xi_2+\Gamma_2\sigma S\right)R_1+\left(\psi_1+\Gamma_1\psi_2\right)R_2.     
 \end{equation}
Note that this result is obtained after recognizing $f(t)=C_1$ or $f(t)=C_2$ in eq. (\ref{HK}) when it corresponds. It has been demonstrated that even in this case of stochastic volatility, it is still possible to create a hedged folio and then at the end we arrive again to the condition (\ref{hedged}), after finding special constraints for $\Gamma_1$ and $\Gamma_2$ such that the white noises are removed. The solution for $\Pi$ is a non-trivial one for this case, and then it requires the definition of the parameter

\begin{eqnarray}   \label{beta}
\beta(S, V, t, r)=\frac{1}{\partial C_1/\partial V}\left(\frac{\partial C_1}{\partial t}+(\lambda+\mu V)\frac{\partial C_1}{\partial S}+\frac{VS^2}{2}\frac{\partial^2 C_1}{\partial S^2}+\rho V^{1/2+\alpha}\zeta\frac{\partial^2 C_1}{\partial S\partial V}+\frac{\zeta^2V^{2\alpha}}{2}\frac{\partial^2 C_1}{\partial V^2}-rC_1\right)\nonumber\\
=\frac{1}{\partial C_2/\partial V}\left(\frac{\partial C_2}{\partial t}+(\lambda+\mu V)\frac{\partial C_2}{\partial S}+\frac{VS^2}{2}\frac{\partial^2 C_2}{\partial S^2}+\rho V^{1/2+\alpha}\zeta\frac{\partial^2 C_2}{\partial S\partial V}+\frac{\zeta^2V^{2\alpha}}{2}\frac{\partial^2 C_2}{\partial V^2}-rC_2\right)
\end{eqnarray}
This parameter does not appear for the case of the BS equation. Indeed $\beta$ in the MG equation is defined as the market price volatility risk because the higher its value is, the lower is the intention of the investors to risk. Take into account that in the MG equation the volatility is a stochastic variable. Since the volatility is not traded in the market, then it is not possible to make a direct hedging process over this quantity \cite{3}. In this way, when we have stochastic volatility, it is necessary to consider the expectations of the investors. This effect appears through the parameter $\beta$. It has been demonstrated in \cite{66} that the value of $\beta$ in agreement with eq. (\ref{beta}) is a non-vanishing result. In general, it is always assumed that the risk of the market (in price) has been included inside the MG equation. The MG equation is then obtained by rewriting the equation (\ref{beta}) in the form

\begin{equation}   \label{MGE}
\frac{\partial C}{\partial t}+rS\frac{\partial C}{\partial S}+(\lambda+\mu V)\frac{\partial C}{\partial V}+\frac{1}{2}VS^2\frac{\partial^2 C}{\partial S^2}+\rho\zeta V^{1/2+\alpha}S\frac{\partial^2 C}{\partial S\partial V}+\zeta^2 V^{2\alpha}\frac{\partial^2 C}{\partial V^2}=rC,   
\end{equation}
where the effects of $\beta$ now appear contained inside the modified parameter $\lambda$ in this equation. In other words, we have shifted the parameter $\lambda\to \lambda-\beta$ in eq. (\ref{MGE}). Later in this paper, we will express this equation in the Hamiltonian form, which is the ideal one for understanding the concept of spontaneous symmetry breaking in Quantum Finance. 

\subsection{Hamiltonian form of the Merton-Garman equation}

The previously analyzed MG equation can be formulated as a Hamiltonian (eigenvalue) equation. We can define a change of variable defined as

\begin{eqnarray}   \label{range}
S=e^x,\;\;\;\;\;-\infty<x<\infty,\nonumber\\
\sigma^2=V=e^y,\;\;\;\;\;-\infty<y<\infty,
\end{eqnarray}
and then the MG equation (\ref{MGE}) becomes \cite{3, 5B, 70}

\begin{equation}
\frac{\partial C}{\partial t}+\left(r-\frac{e^y}{2}\right)\frac{\partial C}{\partial x}+\left(\lambda e^{-y}+\mu-\frac{\zeta^2}{2}e^{2y(\alpha-1)}\right)\frac{\partial C}{\partial y}+\frac{e^y}{2}\frac{\partial^2 C}{\partial x^2}+\rho\zeta e^{y(\alpha-1/2)}\frac{\partial^2 C}{\partial x\partial y}+\zeta^2 e^{2y(\alpha-1)}\frac{\partial^2 C}{\partial y^2}=rC.  
\end{equation}
If we express this equation as an eigenvalue problem in the same form as in eq. (\ref{BSHamiltonian2}) for the BS case, then we have

\begin{equation}
\frac{\partial C}{\partial t}=\hat{H}_{MG}C,   
\end{equation}
with the MG Hamiltonian defined as

\begin{equation}  \label{MGHamilton}
\hat{H}_{MG}=-\frac{e^y}{2}\frac{\partial^2 }{\partial x^2}-\left(r-\frac{e^y}{2}\right)\frac{\partial }{\partial x}-\left(\lambda e^{-y}+\mu-\frac{\zeta^2}{2}e^{2y(\alpha-1)}\right)\frac{\partial}{\partial y}-\rho\zeta e^{y(\alpha-1/2)}\frac{\partial^2 }{\partial x\partial y}-\zeta^2 e^{2y(\alpha-1)}\frac{\partial^2 }{\partial y^2}+r.   
\end{equation}
Exact solutions for the MG equation have been found for the case $\alpha=1$ in \cite{5B} by using path-integral techniques. The same equation has been solved in \cite{45} for the case $\alpha=1/2$ by using standard techniques of differential equations. Note that the equation has two degrees of freedom. Later we will see that when we have spontaneous symmetry breaking, it becomes irrelevant to know the exact solution of this equation.   

\section{The martingale condition in finance}   \label{Sec2}

The martingale condition is required for having a risk-neutral evolution for the price of an Option. This means that the price of a financial instrument is free of any possibility of arbitrage. In probability theory, the risk-free evolution is modeled inside a stochastic process. Assume for example $N+1$ random variables $X_i$, with a joint probability distribution defined as $p(x_1, x_2, ..., x_{N+1})$. Then the martingale process is simply defined as the condition under which

\begin{equation}   \label{martingale}
E[X_{n+1}\vert x_1, x_2, ..., x_n]=x_n,    
\end{equation}
is satisfied \cite{3}. Note that $E[X_i]$ is the expectation value of the random variable. Eq. (\ref{martingale}) suggests that the expected value of a subsequent observation of a random variable is simply the present value. For the purpose of this paper, the random variables correspond to the future prices of the stock given by $S_1$, $S_2$, ..., $S_{N+1}$, which are defined at different times $t_1$, $t_2$, ..., $t_{N+1}$. We can then apply the same martingale condition to the stocks if we make the corresponding discounts in order to compare prices defined at different moments \cite{2, 3}. We can assume that the future value of an equity is defined as $S(t)$. If there is a free-risk evolution of the discounted price defined as

\begin{equation}   \label{securdiscount}
e^{-\int_0^tr(t')dt'}S(t).  
\end{equation}
Then the value follows the martingale process \cite{42}. In this way the conditional probability for the present price is the actual value given by $S(0)$. The martingale condition can then be expressed as \cite{3}

\begin{equation}   \label{interpret}
S(0)=E\left[e^{-\int_0^tr(t')dt'}S(t)\vert S(0)\right],   
\end{equation}
and this result is general. Equivalent expressions have been used for the analysis of the evolution of forward rates \cite{3}. The importance of martingales is analyzed in \cite{80}. The interpretation of eq. (\ref{interpret}) is clear. The left-hand side is just the present price of the security. The right-hand side is the expected value of the discounted price of the security at the time $t$. Discounted means that the quantity evaluated at the time $t$ has to be extrapolated to the present value. Both quantities must be equivalent under the martingale condition.     

\section{The martingale condition as a vacuum condition for a Hamiltonian}   \label{Sec3}

The previous section dealt with the martingale condition. This section deals with the equivalent formulation of the same principle but from the perspective of The Hamiltonian formulation. Consider as before an option on a security $S=e^x$ that matures at time $T$ with the corresponding pay-off function $g(x)$. In this way we can describe the risk-free evolution of the option as 

\begin{equation}   \label{thisone}
C(t, x)= \int_{-\infty}^\infty dx'<x\vert e^{-(T-t)\hat{H}}\vert x'>g(x').    
\end{equation}
By using the previous definition of martingales, for this case we have

\begin{equation}
S(t)=E\left[e^{-(t_*-t)r}S(t_*)\vert S(t)\right].    
\end{equation}
If we introduce $S(x)$ (the price of the security) in eq. (\ref{thisone}), then under the martingale condition, we have 

\begin{equation}
S(t, x)= \int_{-\infty}^\infty dx'<x\vert e^{-(t_*-t)\hat{H}}\vert x'>S(x').    \end{equation}
This equation can be re-expressed in Dirac notation as 

\begin{equation}
<x'\vert S>= \int_{-\infty}^\infty dx'<x\vert e^{-(t_*-t)\hat{H}}\vert x'><x'\vert S>.    
\end{equation}
If we take the base $\vert x'>$ as a complete set of states, then the condition $\hat{I}=\int dx'\vert x'>< x'\vert$ ($\hat{I}$ is the identity matrix) is satisfied and then the previous expression is simplified as

\begin{equation}
\vert S>=e^{-(t_*-t)\hat{H}}\vert S>.    
\end{equation}
Then there is no Hamiltonian (time) evolution for the state $\vert S>$ under the previous conditions. It also comes out that the Hamiltonian annihilates the same state as follows

\begin{equation}   \label{MartHam}
\hat{H}\vert S>=0.    
\end{equation}
Interestingly, the BS Hamiltonian given in eq. (\ref{BSHamiltonian}) as well as the MG Hamiltonian giving in eq. (\ref{MGHamilton}) satisfy the martingale condition in the form defined in eq. (\ref{MartHam}). 

\section{Non-derivative terms introduced in the financial Hamiltonians}   \label{Sec4}

It is possible to introduce potential terms to the BS equation as well as to the MG one as it is explained in \cite{3}. It has been demonstrated that the martingale condition can still be maintained if the potential satisfies some special conditions. In general, a potential term will appear as

\begin{equation}   \label{non-hermBS}
\hat{H}_{BS, MG}^{eff}=\hat{H}_{BS, MG}+\hat{V}(x),    
\end{equation}
with the potential term $\hat{V}$ containing non-derivative terms depending on the security $S$. Since we usually have a change of variables in the Hamiltonian formulation, this functional dependence is indirect. In the previous equation, $\hat{H}_{BH, MG}^{eff}$ is the effective Hamiltonian including the potential contribution. Some barrier options as well as some path-dependent options admit the inclusion of potential terms for their deep understanding \cite{Potential}. On the other hand, for the case of the Black-Scholes Hamiltonian, the martingale condition is maintained if the potential appears in the Hamiltonian in the following form \cite{3}

\begin{equation}   \label{effblack}
\hat{H}_{BS}^{eff}=-\frac{\sigma^2}{2}\frac{\partial^2}{\partial x^2}+\left(\frac{1}{2}\sigma^2-V(x)\right)\frac{\partial}{\partial x}+V(x).    
\end{equation}
Then an effective Hamiltonian expressed in this way can be used for pricing the option. The discount in these general cases depends on the price of the option itself. Then the security discount defined in eq. (\ref{securdiscount}) is modified as \cite{3}

\begin{equation}
e^{-\int_0^tr(t')dt'}S(t)\to e^{-\int_0^tV(x(t'))dt'}S(t).    
\end{equation}
In \cite{3} it is argued that the usual discounting of a security using the spot interest rate $r$ is determined by the argument of no arbitrage involving fixed deposits in the money market account. Studies about viable potentials matching with the reality of the market are under analysis. It is important to notice that the Hamiltonian (\ref{effblack}) can be converted to a Hermitian operator by using a similarity transformation as has been reported in \cite{3}. Then we can define

\begin{equation}   \label{wecandefine}
\hat{H}_{BS}^{eff}=e^s\hat{H}_{Herm}e^{-s}.    
\end{equation}
Here the Hermitian Hamiltonian is defined as

\begin{equation}   \label{Alex}
\hat{H}_{Herm}=-\frac{\sigma^2}{2}\frac{\partial^2}{\partial x^2}+\frac{1}{2}V'(x)+\frac{1}{2\sigma^2}\left(V+\frac{1}{2}\sigma^2\right)^2,    
\end{equation}
and $s=x/2-(1/\sigma^2)\int^x_0 dy V(y)$. This result can be obtained by replacing (\ref{effblack}) and (\ref{Alex}) in eq. (\ref{wecandefine}). From the Hermitian Hamiltonians, it is possible to construct a complete basis and then we can find real eigenvalues associated to this Hamiltonian. Note in particular that in the Black-scholes case $V(x)=r$ is constant. It is a simple task to demonstrate that the Hermitian Hamiltonian obtained by similarity transformation can be also expressed as

\begin{equation}
\hat{H}_{Herm}=e^{\alpha x}\left(-\frac{\sigma^2}{2}\frac{\partial^2}{\partial x^2}+\gamma\right)e^{-\alpha x},  
\end{equation}
with 

\begin{equation}
\gamma=\frac{1}{2\sigma^2}\left(r+\frac{1}{2}\sigma^2\right)^2\;\;\;\;\;\alpha=\frac{1}{\sigma^2}\left(\frac{1}{2}\sigma^2-r\right).    
\end{equation}
Among the trivial examples of potentials already analyzed in the literature, we find the Down-and-Out barrier option, where the stock price has to be over some minimal value, below which it becomes worthless. This behavior can be guarantee with an infinite potential barrier boundary condition imposed for the value of the corresponding price. This case can be worked out directly from the non-Hermitian Hamiltonian defined in eq. (\ref{non-hermBS}). Another example of potential corresponds to the Double-Knock-Out barrier option, where it is easier to work with the Hermitian Hamiltonian part as it is defined in eq. (\ref{wecandefine}) but including the potential part as follows

\begin{equation}   \label{standarddefinitio}
\hat{H}_{DB}=\hat{H}_{BS}+\hat{V}(x)=e^s\left(\hat{H}_{Herm}+\hat{V}(x)\right)e^{-s}.    
\end{equation}
This definition is used for analyzing cases where the stock has to be maintained between a maximal and a minimal value. Note that the definition (\ref{standarddefinitio}) can be also used for the analysis of the Down-and-Out barrier if we focus on the non-Hermitian part (the term in the middle of the equation) with the corresponding potential. Note that these examples of potentials representing real situations are trivial cases. More details about these examples can be found in \cite{3}.   

\section{Deeper analysis for the Black-Scholes and the Merton-Garman equation}   \label{NS1}

If we analyze the martingale condition, we can notice that its interpretation as a vacuum condition is not perfect. The reason is that although the Hamiltonian annihilates the martingale state $\vert S>$, as can be seen from eq. (\ref{MartHam}), the momentum operator corresponding to the prices of the options ($\hat{p}_x$) does not annihilate the vacuum perfectly. This can be seen from their definitions as follows 

\begin{equation}
\hat{p}_x\vert S>=e^x\vert S>.    
\end{equation}
This means that the symmetry under translations of the prices, carried out from the security $S$ is spontaneously broken. An exception is the case where $S\to0$ and then $x\to-\infty$ as can be seen from eq. (\ref{range}). This however, would give us a trivial value for the security as $S=0$. For general values, the symmetry under translations of prices is spontaneously broken. This means that the different values of $S$ represent different possible vacuums. The action of $\hat{p}_x$ over $\vert S>$, simply maps one vacuum toward another one defined through the selected value for $x$. Such operation could be seen as a rotation in a complex plane if we make the transformation of variables $x\to i n\theta$. In such a case, we have 

\begin{equation}
\hat{p}_x\vert S>=e^{in\theta}\vert S>=\vert S'>\neq\vert S>. 
\end{equation}
Here $n$ is just a constant for helping the phase $\theta$ to be dimensionless. After this change of variable, the action of $\hat{p}_x$ is to map one vacuum into another one through a rotation defined by the phase $\theta$. This condition shows the vacuum degeneracy. The same condition has different meanings depending on whether we consider the BS or the MG equation. This important detail about the Martingale condition deserves more attention. 

\subsection{Reinterpretation of the Martingale condition}                                                           
The martingale condition can be reinterpreted if we make the following changes of variable in eqns. (\ref{BSHamiltonian}) and (\ref{MGHamilton})

\begin{equation}   \label{specialcond}
\hat{p}C(x, t)=-i\frac{\partial C(x, t)}{\partial x},\;\;\;\;\;i\hat{p}C(x, t)=\hat{\phi}.   
\end{equation}
In this way we convert the derivative field $\partial C(x, t)/\partial x$ into a non-derivative field. Then the BS Hamiltonian can be expressed as

\begin{equation}
\hat{H}_{BS}C(x, t)=-\frac{\sigma^2}{2}\hat{\phi}^2+\left(\frac{1}{2}\sigma^2-r\right)\hat{\phi}+rC(x, t).    
\end{equation}
There are no kinetic terms in this Hamiltonian when expressed in terms of the field $\hat{\phi}$. Then we can find the vacuum condition for this case as $\partial \hat{H}_{BS}/\partial \hat{\phi}=0$, obtaining then 

\begin{equation}
\hat{\phi}_{vac}=\left(\frac{r}{\sigma^2}-\frac{1}{2}\right).
\end{equation}
We can associate this result to the martingale condition. We can notice that the momentum only annihilates the vacuum when $r=\frac{\sigma^2}{2}$. This is the same condition under which the BS Hamiltonian becomes Hermitian. From this we conclude that the vacuum is single (annihilated by the momentum operator) when the Hamiltonian is Hermitian, or equivalently, when there is no flow of information through the boundaries of the system. This is a remarkable property of the BS equation. On the other hand, when there is flow of information through the boundary of the system, then the momentum cannot annihilate the vacuum and then the martingale condition is degenerate. This is in fact spontaneous symmetry breaking from the perspective of the BS equation. Note that the martingale condition requires additionally

\begin{equation}   \label{multime}
S(x, t)=\hat{\phi}_{vac}^n(x, t),\;\;\;\;\;n=1, 2,3,...n.    
\end{equation}
This is an additional constraint for the martingale condition which guarantees that $S(x, t)=\hat{\phi}_{vac}=\left(\frac{r}{\sigma^2}-\frac{1}{2}\right)$. It is interesting to notice that the spontaneous symmetry breaking under the changes of the prices of the options (the fact that the momentum associated to this symmetry cannot annihilate the ground state), happens when there is flow of information through the boundaries of the system. This previous result requires the additional constraint

\begin{equation}
S(x, t)=e^x=\left(\frac{r}{\sigma^2}-\frac{1}{2}\right).    
\end{equation}
This means that the no-flow of information only happens when the prices in the market vanish trivially under the vacuum condition.  

\subsection{The spontaneous symmetry breaking for the Merton Garman case}

Following the same pattern just explained for the BS equation, we can analyze the MG equation in a similar fashion, by using the special condition (\ref{specialcond}). Under this approximation, the Martingale condition, taken as the ground state for the field $\hat{\phi}$, is given by

\begin{equation}   \label{Thisoneyyy}
\hat{\phi}_{vac}=\left(\frac{r}{e^y}-\frac{1}{2}\right).    
\end{equation}
The previous arguments are still valid with the difference that the volatility is stochastic in this case. Once again the vacuum condition vanishes when there is no flow of information through the boundaries of the system. Then when the Hamiltonian is Hermitian, the martingale condition is unique, otherwise, it becomes degenerate and the symmetry under changes of the prices of the Options is spontaneously broken. The single vacuum condition here becomes

\begin{equation}   \label{ybebe}
r=\frac{e^y}{2}.    
\end{equation}
The symmetry is spontaneously broken when $r\neq \frac{e^y}{2}$. Still the condition (\ref{multime}) is valid here.

\subsection{A more general condition for the symmetry breaking in the Merton Garman equation}

When we analyze the MG equation, the martingale condition is normally taken such that it is independent on the stochastic volatility. The volatility is a function of the variable $y$, as it is defined in eq. (\ref{range}). If the martingale condition is taken as independent of $y$, then any term with derivative with respect to this variable, will annihilate the state (\ref{Thisoneyyy}). From this perspective, $y$ can be taken as fixed when we are determining the vacuum conditions. Here we would like to define a more general martingale condition, such that the possible changes in $y$ can be considered. We will take the martingale state as

\begin{equation}   \label{newmartin}
\hat{H}_{MG}e^{x+y}=\hat{H}_{MG}S(x, y, t)=0.    
\end{equation}
Here we extend the arguments showed in (\ref{range}), considering the extensions of the original martingale state $S(x, t)=e^x$. The condition (\ref{newmartin}) will be considered here as the extended martingales condition with $S_{NM}(x, y, t)=e^{x+y}$. By using the Hamiltonian (\ref{MGHamilton}) and the result (\ref{newmartin}), we can obtain the condition for the Hamiltonian to annihilate the martingale state as

\begin{equation}   \label{corona4}
\lambda+e^y\left(\mu+\frac{\zeta^2}{2}e^{2y(\alpha-1)}+\rho\zeta e^{y(\alpha-1/2)}\right)=0, 
\end{equation}
if $e^x\neq0$, avoiding then a non-trivial result. This previous condition is necessary for the state $e^{x+y}$ to be considered as the martingale state.  

\subsubsection{The extended martingale condition and the flow of information}

Previously, when we considered the ordinary martingale condition, we could demonstrate that it can be also considered as a vacuum state. The vacuum is single if there is no flow of information through the boundaries of the system and it is degenerate if the information flows through the boundaries of the system. When the vacuum is single, the momentum, defined as the generator of the changes in prices, is a perfect symmetry. On the other hand, when the vacuum is degenerate, then the same symmetry is spontaneously broken because although the Hamiltonian annihilates the ground state (martingale condition), the momentum does not do it. The interesting point here is the connection between spontaneous symmetry breaking and flow of information. Something which has been suggested before in \cite{Winchi} in a different context. Repeating some of our previous arguments, we can promote the action of the momentum operator over the state $C(x, y, t)$ to be a Quantum field. In the case of the MG equation, we have two of such fields, here defined as

\begin{equation}
i\hat{p}_xC(x, y, t)=\hat{\phi}_x,\;\;\;\;\;i\hat{p}_yC(x, y, t)=\hat{\phi}_y.    
\end{equation}
Here we have defined

\begin{equation}
\hat{p}_xC(x, y, t)=-i\frac{\partial}{\partial x}C(x, y, t),\;\;\;\;\;\hat{p}_yC(x, y, t)=-i\frac{\partial}{\partial y}C(x, y, t). \end{equation}
Under these definitions, in the neighborhood of the vacuum state, taken as the martingale state, we can express the action of the MG Hamiltonian as

\begin{equation}
\hat{H}_{MG}C(x, y, t)=-\frac{e^y}{2}\hat{\phi}^2_x-\left(r-\frac{e^y}{2}\right)\hat{\phi}_x-\left(\lambda e^{-y}+\mu-\frac{\zeta^2}{2}e^{2y(\alpha-1)}\right)\hat{\phi}_y-\rho\zeta e^{y(\alpha-1/2)}\hat{\phi}_x\hat{\phi_y}-\zeta^2e^{2y(\alpha-1)}\hat{\phi}^2_y+rC(x, y, t).    
\end{equation}
This result is based on the MG Hamiltonian defined in eq. (\ref{MGHamilton}). We can find the extremal condition for this Hamiltonian by calculating $\frac{\partial \hat{H}_{MG}}{\partial \hat{\phi}_x}$ and $\frac{\partial \hat{H}_{MG}}{\partial \hat{\phi}_y}$. The vacuum conditions are then determined by the following set of equations

\begin{equation}   \label{corona1}
e^y\hat{\phi}_{x vac}+\left(r-\frac{e^y}{2}\right)+\rho\zeta e^{y(\alpha-1/2)}\hat{\phi}_{y vac}=0.
\end{equation}
\begin{equation}   \label{corona2}
\lambda e^{-y}+\mu-\frac{\zeta^2}{2}e^{2y(\alpha-1)}+\rho\zeta e^{y(\alpha-1/2)}\hat{\phi}_{x vac}+2\zeta^2e^{2y(\alpha-1)}\hat{\phi}_{y vac}=0.    
\end{equation}
This system of equations can be easily solved, showing then that in general $\hat{\phi}_{x vac}\neq0$ and $\hat{\phi}_{y vac}\neq0$. This result only reflects the fact that the momentum operators $\hat{p}_x$ and $\hat{p}_y$ do not annihilate the vacuum state taken as the martingale state in general. Then both symmetries, namely, the symmetries under changes of the prices and the symmetries under changes of the stochastic volatility are spontaneously broken in this particular situation. Note that here again, following the standard conditions for the martingale state, we have taken 

\begin{equation}
S(x, y, t)=X(x)Y(y)=\hat{\phi}_{x vac}^n\hat{\phi}_{y vac}^m, \;\;\;\;\;n=1, 2, 3, ...\;\;\;\;\;m=1, 2, 3, ...
\end{equation}
We can still find here the conditions under which the vacuum state is annihilated by the momentum, or equivalently, the conditions when $\hat{\phi}_{x vac}=\hat{\phi}_{y vac}=0$. If we take the no-flow of information condition through the boundaries of the system, the conditions (\ref{corona1}) and (\ref{corona2}) are simplified to

\begin{eqnarray}
e^y\hat{\phi}_{x vac}=-\rho\zeta e^{y(\alpha-1/2)}\hat{\phi}_{y vac}, \nonumber\\
\rho\zeta e^{y(\alpha-1/2)}\hat{\phi}_{x vac}=-2\zeta^2e^{2y(\alpha-1)}\hat{\phi}_{y vac}.
\end{eqnarray}
Note that the only way how this vacuum state vanishes, restoring then the broken symmetry, is is to have $\rho=0$. If we go to the basic concepts of the MG equation, this would mean zero correlation between the white noises for the prices and volatility. In other words $<R_1R_2>=0$ in agreement with the definitions given in eqns. (\ref{corona3}). This is a remarkable result and it complements the no-flow of information condition for this case. Note however, that one more condition is needed in order to satisfy the martingale condition for the state $S(x, y, t)=e^{x+y}=\hat{\phi}_{x vac}\hat{\phi}_{y vac}$. This extra condition is

\begin{equation}
\zeta\to0,\;\;\;\;\;with\;\;\;\;\;\frac{\rho}{\zeta}\to0.    
\end{equation}
This condition comes if we replace 

\begin{equation}   \label{corona6}
r=\frac{e^y}{2},\;\;\;\;\;\lambda e^{-y}+\mu-\frac{\zeta^2}{2}e^{2y(\alpha-1)}=0,    
\end{equation}
together with $\rho=0$ in eq. (\ref{corona4}). $\zeta=0$ is a natural consequence of this calculation. The extra condition $\rho/\zeta\to0$ is necessary for getting $\hat{\phi}_{y vac}=0$. $\zeta=0$ means total absence of white noise coming from the stochastic volatility as can be seen from the derivations of the MG equation in eqns. (\ref{corona5}). Note that the conditions (\ref{corona6}) guarantee the Hamiltonian to be Hermitian and then this guarantees the no-flow of information through the boundaries of the system. It is interesting to notice that the white noise coming from the fluctuations on the prices does not have to vanish for the vacuum state to be single. This means that the white noise for the prices might still remain finite under such special circumstances.  

\section{Additional broken symmetries}   \label{NS2}   

In more general situations, involving other symmetries for the system, some potential terms might appear in both equations, namely, the BS and the MG equations. It is evident from their structure, that the equations (\ref{BSHamiltonian}) and (\ref{MGE}), are both invariant under any transformation keeping $x$ constant. This is equivalent to transformations able to keep the price of the stock fixed. Those transformations might imply some flow of information but keeping invariant the price for the case of the BS and the MG equations as well as the volatility for the case of the MG equation alone. From the geometrical point of view, the new symmetries can be interpreted as a rotation over a plane where the price of the Stock $S$ is fixed. The generator of the new symmetries, can be represented conveniently as an operator 

\begin{equation}
\hat{L}=-i(\bf {r}\times{\bf \nabla}).    
\end{equation}
Here ${\bf \nabla}$ is reduced to a derivative with respect to $x$, which is related to the price of the stock. We can then define a unitary operator which reproduces the rotations of the system keeping $x$ unchanged. It is $\hat{U}=e^{-i\hat{\bf L}\theta}$. Here $\theta$ is the "angle" of rotation. This rotation can be interpreted from the perspective of finance as a change of the conditions of the market such that the price of the stock keeps the same anyway. These mentioned changes in the conditions of the market, might include a corresponding flow of information, but keeping the prices unchanged. Note that this is different to the symmetry transformations generated by the momentum $\hat{p}_x$ and $\hat{p}_y$, illustrated before in the standard scenarios of the BS and the MG equations.
In standard conditions, if the price does not change, then we can summarize the effect of the operator $\hat{U}$ as

\begin{equation}
\hat{U}\hat{H}C(x, t)=\hat{H}\hat{U}C(x, t)=0.    
\end{equation}
This also implies that the operator $\hat{L}$ is a conserved quantity analogous to the angular momentum in ordinary Quantum Mechanics \cite{angular}. We can also notice that the conservation of $\hat{L}$ is equivalent to the condition

\begin{equation}
[\hat{U}, \hat{H}]=0.    
\end{equation}
We are particularly interested in the cases where the martingale condition $\hat{H}\vert S>=0$ is satisfied but in addition, we have

\begin{equation}
\hat{U}\vert S>\neq\vert S>.    
\end{equation}
This means that even if the martingale condition is satisfied, and even if the Hamiltonian is invariant under the transformations defined by the operator $\hat{U}$ ($[\hat{U}, \hat{H}]=0$), still the operator $\hat{L}$ does not annihilate the state $\vert S>$, even if $\hat{L}$ represents a symmetry of the Hamiltonian itself. Here we interpret this phenomena as a multiplicity of martingale conditions. One example of this situation corresponds to the following Hamiltonian

\begin{equation}   \label{totality}
\hat{H}=\hat{H}_{BS, MG}+\hat{V}(x, t).    
\end{equation}
If we take $\psi(S, t)=<x\vert S>$ as a field, then we can introduce a potential term as follows   

\begin{equation}   \label{Potentialspontaneous}
V(x, t)=-\mu^2C^2(x, t)+\omega C^4(x, t),    
\end{equation}
which in operator notation would be 

\begin{equation}   \label{usedequation}
\hat{V}(x, t)\vert S>=V(x, t),\;\;\;\;\;\; \hat{V}(x, t)=-\mu^2(<x\vert)^2+\omega (<x\vert)^4.
\end{equation}
Note that the action of the operator $\hat{x}$ is defined as $\hat{x}\vert S>=x\vert S>$. The martingale condition is satisfied in the neighborhood of the minimal of the potential (\ref{usedequation}). This minimal of the potential marks the zone where we can neglect all the kinetic terms in the MG and the BS equations.Then, for both equations, the MG and the BS, around the minimal of the potential (\ref{Potentialspontaneous}), we get 

\begin{equation}   \label{consition1}
\hat{H}_{BS}S(x, t)\approx \hat{H}_{MG}S(x, t)=0,    
\end{equation}
and then the total Hamiltonian for both cases is in agreement with eq. (\ref{totality}) 

\begin{equation}   \label{condition2}
\hat{H}S(x, t)\approx\hat{V}S(x, t)=-\mu^2S^2(x, t)+\omega S^4(x, t).    
\end{equation} 
Here we have used the result (\ref{usedequation}). The conditions (\ref{consition1}) and (\ref{condition2}) are achieved around the minimal 

\begin{equation}   \label{superman}
\frac{\partial V}{\partial S(x,t)}=0.    
\end{equation}
If the term with the coefficient $\mu^2$ is dominant, then we have a single vacuum condition given by $S(x, t)=0$ and the conclusions are trivial. However, if the term with the coefficient $\omega$ is the dominant one, then the vacuum condition is non-trivial and it is given by

\begin{equation}   \label{minimal}
\vert S(x, y)\vert=\left\vert\frac{1}{\sqrt{2}\omega}\mu\right\vert.    
\end{equation}
This fixes the magnitude of the vacuum state, but it does not fix its direction. Since the direction of the field $S(x, t)$ is arbitrary, then we can conclude that we have a multiplicity of possible martingale conditions. Then our market will be in equilibrium no matter which state (\ref{minimal}) is selected. These conditions will have the same price. This situation is different to the symmetries analyzed before where the prices were allowed to change. Here the analysis is reduced to the states satisfying the vacuum condition (\ref{superman}). Under these conditions, there is no distinction between the different financial equations if they only contain derivative terms. This is the case of the MG and BS equations. Around the minimal, all the derivative terms are negligible, having then no distinction between MG or BS equation. In this way we have then found a universal behavior for the financial market. One consequence of this result, is that the discounted price defined in eq. (\ref{securdiscount}), might consists on many different possible paths, depending on which vacuum direction is selected with the magnitude defined eq. (\ref{minimal}). Each vacuum represents an equilibrium condition of the market. But each equilibrium condition, although equivalent, corresponds to different configurations of the system. This means that it is possible to get the same prices in the market under different conditions. For example, a presidential decision might have the same effect on the price of an option as a protest. Then if we assume that the market reach an equilibrium after a presidential decision, the price of the stock might be the same as the equilibrium condition obtained after a protest (without any presidential decision). Note that although both events would correspond to different Hilbert spaces, they would generate the same prices over an option because they are connected to the same martingale condition. It is possible to arrive to one condition from the other by applying the broken symmetry operator $\hat{U}$ to the vacuum as many times as it is necessary for changing from one configuration to the other.         

\section{Conclusions}   \label{Sec5}

In this paper we have demonstrated that the martingale condition can be perceived as a vacuum condition when we express the Financial equations in the Hamiltonian form. We have demonstrated that even if the Hamiltonian annihilates this condition, the symmetries under changes of the prices of the Options cannot annihilate the vacuum (taken as the martingale state). These symmetries are then spontaneously broken for the BS as well as for the MG equations. We found the conditions where the vacuum, in these two situations, becomes unique. This happens when there is no flow of information through the boundaries if the system, or equivalently, the corresponding Hamiltonians become Hermitian.
We could extend the martingale condition for including the symmetries under changes of the stochastic volatility. We defined then the momentum for the volatility as the generator of these transformations. We found the conditions under which the martingale state (vacuum) becomes unique. This happens when: 1). There is no flow of information through the boundaries of the system. 2). There is no correlation between the white noise corresponding to the prices with the white noise corresponding to the volatility. There is still one additional condition is necessary for satisfying the martingale condition in this extended sense and it corresponds to 3). The complete absence of the white noise corresponding to the stochastic volatility. In order to complete our results, we added some potential terms to the BS as well as for the MG equation and we analyzed the situations where the symmetries under "rotations", or equivalently, symmetries keeping the prices unchanged ($x$ constant), are spontaneously broken. In these special situations, the BS and the MG equations behave identically because all the derivative terms are negligible around the vacuum (martingale state). These cases correspond to a multiplicity of martingale states, all of them with the same price but different market conditions. This is the first paper able to define in a clear and consistent form, the concept of spontaneously symmetry breaking for the stock markets based on some financial equations.\\\\

{\bf Acknowledgement}
I. A. would like to thank to Prof. Yu Chen from the University of Tokyo for providing useful references and bibliography, as well as very useful discussions. The authors also would like to thank the Institute of International Business and Governance of the Open University of Hong Kong, partially funded by a grant from the Research Grants Council of the Hong Kong Special Administrative Region, China (UGC/IDS16/17), for its support.

\bibliographystyle{unsrt}  


\begin{thebibliography}{1}

\bibitem{1} N. J. Balsara, "Money Management Strategies for Futures Traders", Wiley Finance, (1992); R. Mansuy, "The origins of the Word "Martingale", Electronic Journal for History of Probability and Statistics. 5 (1), (2009).  
\bibitem{2} J. Y. Campbell, A. S. Low and A. C. Mackinlay, "The Econometrics of Financial Markets", Princeton University Press (1997); J. M. Harrison and S. Pliska, "Martingales and Stochastic Integrals in the Theory of Continuous Trading", Stochastic Processes and their Applications, 11 (1981): 215; S. Sundaresan, "Fixed Income Markets and their Derivatives", South-Western College Publishing (1997). 
\bibitem{3}  B. E. Baaquie, ”Quantum Finance: Path integrals and Hamiltonians for options and interestrates”, Cambridge University Press (2004), pp 52-75. 
\bibitem{4} L. Haughton (ed.), "Vasicek and Beyond Risk Publications" (1994).
\bibitem{7} Y. Nambu and G. Jona-Lasinio, "Dynamical Model of Elementary Particles Based on an Analogy with Superconductivity I". Phys. Rev. 1961, {\bf 122}, 345; Nambu, Y.; G. Jona-Lasinio, "Dynamical model of elementary particles based on an analogy with superconductivity II". Phys. Rev. 1961, {\bf 124}, 246.
\bibitem{Winchi} S. J. Gu, W.C. Yu and H.Q. Lin, "Construct order parameter from the spectra of mutual information",
Annals of Physics {\bf 336}, 118-129; W. C. Yu, S. J. Gu and H. Q. Lin, "Density matrix spectra and order parameters in the 1D extended Hubbard model", EPJB,  {\bf 89} (10), 212; W. C. Yu, Y. C. Li, P. D. Sacramento and H. Q. Lin, "Reduced density matrix and order parameters of a topological insulator", Phys. Rev. B {\bf 94} (24), 245123.
\bibitem{5} V. Linetsky, "The Path Integral Approach to Financial Modeling and Options Pricing”, Computational Economics 11 (1998) 129; Mathematical Finance 3 349, (1993); B. E. Baaquie, C. Coriano and M. Srikant, "Hamiltonian and Potentials in Derivative Pricing Models: Exact Results and Lattice Simulations", Physica A 334 (3) (2004): 531–57; http://xxx.lanl.gov/cond-mat/0211489.
\bibitem{6} I. Arraut. A. Au, A. Ching-Biu Tse and C. Segovia, "The connection between multiple prices of an Option at a given time with single prices defined at different times: The concept of weak-value in quantum finance", Physica A {\bf 526} (2019) 121028.  
\bibitem{9} F. Black and M. Scholes, "The pricing of options and corporate liabilities", J. Pol. Ec., 81 (1973), p. 637.
\bibitem{Europe} J. C. Hull, {\it Options, futures and other derivatives}, Fifth Edition, Prentice-Hall International (2003).
\bibitem{Epjons} E. P. Jones, {\it Option arbitrage and strategy with large price changes}, Journ. Fin. Econ, {\bf 13}, (1984):91.
\bibitem{Op2} F. Black and M. Scholes, {\it The pricing of Options and Corporate Liabilities}, Journ. Pol. Ec. {\bf 81} (1973):637.
\bibitem{Op3} R. C. Merton, {\it The theory of rational Option Pricing}, Bell Journ. Econ. Manag. Sc. {\bf 4} (1973):141-183.
\bibitem{Merton2} R. C. Merton, {\it Option pricing when underlying stock returns are discontinous}, Journ. Fin. Econ. {\bf 3}, (1976):125.
\bibitem{8} J. Alexandre, J. Ellis, P. Millington and D. Seynaeve, "Gauge invariance and the Englert-Brout-Higgs mechanism in non-Hermitian field theories", Phys.Rev. D99 (2019) no.7, 075024;  J. Alexandre, J. Ellis, P. Millington and D. Seynaeve, "Spontaneous symmetry breaking and the Goldstone theorem in non-Hermitian field theories", Phys.Rev. D98 (2018) 045001;  J. Alexandre, P. Millington and D. Seynaeve, "Symmetries and conservation laws in non-Hermitian field theories", Phys.Rev. D96 (2017) no.6, 065027.
\bibitem{45} S. L. Heston, {\it A Closed-Form Solution for Options with Stochastic Volatility with Application to Bond and Currency Options}, The Review of Financial Studies, 6 (1993): 327; J. C. Hull and A. White, {\it An Analysis of the Bias in Option Pricing Caused by a
Stochastic Volatility}, Advances in Futures and Options Research, 3 (1988): 27; J. C. Hull and A. White, {\it The Pricing of Options on Assets with Stochastic Volatilities}, The Journal of Finance, 42 (2) (June 1987): 281; H. Johnson and D. Shanno, {\it Option Pricing when the Variance is Changing}, Journal of Financial and Quantitative Analysis, 22 143; L. H. Mervill and D. R. Pieptea, {\it Stock Price Volatility: Mean-Reverting Diffusion and Noise}, Journal of Financial Economics, 24 (1989): 193; J. M. Poterba and L. H. Summers, {\it The Persistence of Volatility and Stock Market Fluctuations}, American Economic Review, 76 (1986): 1142; L. O. Scott, {\it Option Pricing When the Variance Changes Randomly: Theory, Estimation and an Application}, Journal of Financial and Quantitative Analysis, 22 (1987): 419.
\bibitem{51} J. C. Hull, {\it Options, Futures and Other Derivatives}, Fifth Edition, Prentice-Hall International (2003).
\bibitem{66} C. G. Lamoureux and W. D. Lastrapes, {\it Forecasting Stock-Return Variance: Toward an Understanding of Stochastic Implied Volatilities}, Review of Financial Studies, 6 (1993): 293.
\bibitem{5B} B. E. Baaquie, {\it A Path Integral Approach to Option Pricing with Stochastic Volatility: Some Exact Results}, Journal de Physique I, 7 (12) (1997) 1733.
\bibitem{70} S. Marakani, {\it Option Pricing with Stochastic Volatility}, Honours Thesis. National University of Singapore (1998).
\bibitem{42} L. Haughton (ed.), {\it Vasicek and Beyond}, Risk Publications (1994).
\bibitem{80} M. Musiela and M. Rutkowski, {\it Martingale Methods in Financial Modeling}, Berlin, Springer (1997).
\bibitem{Potential} V. Linetsky, {\it The Path Integral Approach to Financial Modeling and Options Pricing}, Computational Economics 11 (1998) 129; Mathematical Finance 3 349 (1993); B. E. Baaquie, C. Coriano, and M. Srikant, {\it Hamiltonian and Potentials in Derivative Pricing Models: Exact Results and Lattice Simulations}, Physica A 334 (3) (2004): 531–57; http://xxx.lanl.gov/cond-mat/0211489.
\bibitem{angular} D. J. Griffiths, {\it Introduction to Quantum Mechanics}, Prentice Hall. p. 146, (1995).

\end{thebibliography}

\end{document}